\newcommand{\diracslash}[1]{#1\llap{/\kern2pt}}

\newcommand{\be}{\begin{equation}}
\newcommand{\ee}{\end{equation}}
\newcommand{\bea}{\begin{eqnarray}}
\newcommand{\eea}{\end{eqnarray}}
\newcommand{\ba}[1]{\begin{array}{#1}}
\newcommand{\ea}{\end{array}}

\newcommand{\bt}{\begin{tabular}}
\newcommand{\et}{\end{tabular}}

\newcommand{\beas}{\begin{eqnarray*}}
\newcommand{\eeas}{\end{eqnarray*}}

\documentclass[preprint,prd,aps,floats,nofootinbib,floatfix]{revtex4}

\DeclareSymbolFont{rsfs}{U}{rsfs}{m}{n}
\DeclareSymbolFontAlphabet{\mathrsfs}{rsfs}
\usepackage{graphicx}
\usepackage{multirow}

\usepackage{aliascnt}

\usepackage{pgf}

\usepackage{pdfpages}
\usepackage{amsmath}
\usepackage[amsmath,thmmarks]{ntheorem}
\usepackage{csquotes}

\usepackage{tabularx}
\usepackage{graphicx}
\usepackage{cleveref}

\begin{document}


\title{$\psi$(4040) and $Y(4008)$ decay widths in hot and dense strange hadronic matter}

\author{Rahul Chhabra}
\email{rahulchhabra@ymail.com,}
\affiliation{Department of Physics, Dr. B R Ambedkar National Institute of Technology Jalandhar,
 Jalandhar -- 144011,Punjab, India}


\def\be{\begin{equation}}
\def\ee{\end{equation}}
\def\bearr{\begin{eqnarray}}
\def\eearr{\end{eqnarray}}
\def\zbf#1{{\bf {#1}}}
\def\bfm#1{\mbox{\boldmath $#1$}}
\def\hf{\frac{1}{2}}
\def\kp{\zbf k+\frac{\zbf q}{2}}
\def\km{-\zbf k+\frac{\zbf q}{2}}
\def\hwo{\hat\omega_1}
\def\hwt{\hat\omega_2}

\begin{abstract}

In the present investigation, we evaluate the in-medium partial decay widths of exited states $\psi$(4040) and $Y$(4008) decaying to  the  pairs of non-strange pseudoscalar $D \bar{D}$ mesons, strange pseudoscalar $D_s \bar{D}_s$ mesons, non-strange vector $D^* \bar{D}^*$ meson and pseudoscalar-vector $D \bar{D}^*$ mesons using $^3P_0$ model. The in-medium effects are incorporated through the in-medium masses of daughter mesons (calculated using the chiral SU(3) model and QCD sum rule approach in our previous works). We consider $\psi$(4040) and $Y$(4008) states as  3$^3S_1$ states and observe the in-medium dominance of one state over the other for a given decay mode.   The results of the present investigation will prove as one step forward in assigning the correct spectroscopic state to controversial $Y$(4008) state.

%
\end{abstract}

\maketitle
\section{Introduction}

In past few years, many hidden charmed states have been observed \cite{y2,y3,y4,y5,y6,y7,y8,y9,y10,y11,y12,y13,y14}, which encourage us to enhance the knowledge of the non-perturbative regime of QCD. Many of these states have been confirmed by the theoretical and experimental works. But their are some states which still needed to be confirmed. For example, a state with mass 4008 $\pm$ 40$^{+114} _{-28}$ MeV and decay width $\Gamma$ = 226$\pm$44$\pm$87 MeV, is still controversial. This was first observed by the Belle collaboration \cite{belle}. Soon after this,  $Y(4008)$ state was not confirmed in the Babar collaboration. After a while updated version of the Belle collaboration again confirmed this state \cite{upbelle}. This inconsistency continues with the fresh results of BESII collaboration where the state $Y(4008)$ was not found \cite{besiii}.   In-order to have clear understanding of a particular state great attention from both the theoretical as well as experimental works is required. In this respect, authors tried to explain the properties of $Y(4008)$ by considering it as $D^* \bar{D^*}$ molecular state as well as $\psi(3S)$ charmonium state \cite{dmol}. In this work, author observed comparable branching ratios of $Y(4008) \to$  $J/\psi \pi^+ \pi^-$ and $Y(4008) \to$  $J/\psi \pi^0 \pi^0$ and experimental search for the new hidden channels like $D\bar{D}$, $D\bar{D^*}$ were proposed. Furthermore, using  non-relativistic screened potential model, author proposed $Y(4008)$ as $\psi(3S)$ state \cite{screen}. Also, using $^3 P_0$ model, authors observed strong decay widths of $Y(4008)$ state by considering it as $\psi(3^3S_1)$ state \cite{is}. In this work authors predicted that, $Y(4008)$ state can be confirmed as $\psi(3^3S_1)$ state once it is observed above $D^* \bar{D^*}$ threshold. 
Apart from this, in ref. \cite{dmol2}  $Y(4008)$ state was investigated as $D^* \bar{D^*}$ molecule using one boson exchange model but this idea was further rejected because of the huge width of the state. Furthermore, in ref. \cite{jp}, using coulomb gauge Hamiltonian approach,  $Y(4008)$ state was proposed as lightest molecule of $1^{--}$ type with $\eta h_c$ rather than $D \bar{D}$ structure. However, using flux tube model with four confinement model authors interpreted  $Y(4008)$ state as tetraquark state [cq][$\bar{c}\bar{q}$] with $^1 P_1$. Moreover, in ref. \cite{res} authors applied unified Fano-like interference picture to electron annihilation process $e^+ e^- \to \pi^+ {\pi^-} J/\psi$, and argued that $Y(4008)$ is not a genuine resonance. Recently, authors studied $\psi(4040)$ state resonance using QFT approach \cite{pio}. In this approach, authors argued that $Y$(4008) may be treated as a peak generated by $\psi$(4040) and $\bar{D^*}D^*$ loops with $\pi^+ \pi^- J/\psi$ in the final state \cite{pio}.  Here, in  all the above discussed calculations  no medium effects were included. 

In future the PANDA experiment  that will  be benefited from the high intense $\bar{p}$ beams which will be provided by the HESR of FAIR (in the momentum range 1.5 $-$ 15 GeV) is expected to explore wide range of charmonium states \cite{panda1,panda2,panda3}. Hence there is a finite possibility of production of these charmonium states in $p\bar{p}$ annihilation of PANDA experiment. Therefore, it is important to have prior knowledge of the properties of $Y(4008)$ and $\psi$(4040) states and the possible impact of the presence of the other mesons in the medium. In this way,  in the present investigation, we observed the in-medium strong decays of the states $\psi(4040)$/$Y(4008)$ decaying to  $D\bar{D}$, $D\bar{D^*}$, $D^* \bar{D^*}$ and $D_s \bar{D}_s$ pairs, using $^3 P_0$ model and tried to put forward some possible conclusions to predict $Y(4008)$ state in future. Further, the state $Y(4008)$ lies much closer to $\psi$(4040) state, which is well known state  with quantum number $\psi(3^3S_1)$, therefore in the present work we considered the same state  for $Y(4008)$.
The medium effects are incorporated through the in-medium masses of $D^*$, $D$ and $D_s$ mesons calculated in our previous works \cite{rahul1,rahul2,rahul3}. In these papers, by using chiral SU(3) model along with QCD sum rule approach we investigated the in-medium masses of vector $D^*$ ($D^{*+} D^{*0}$) \cite{rahul1},  pseudoscalar $D$ $(D^+ D^0)$ \cite{rahul2} and strange pseudoscalar $D_s$ \cite{rahul3} mesons in hot and dense strange asymmetric medium.

 The outline of the paper is as follows:
In section II we first give brief discussion of chiral SU(3) model and QCD sum rule approach used to calculate the in-medium mass of open charm mesons. Section III will be devoted to the $^3 P_0$ model used to evaluate in-medium strong decays of $Y(4008)$ and $\psi$(4040) states. Then this will be followed by results and discussion in section IV. We will summarise our work in section V.

 
\section{Chiral SU(3) model and QCD sum rule approach}

Chiral SU(3) model is an effective theory based on the chiral property of quarks  and non-realization of chiral symmetry. In this model, we start with the effective Lagrangian density which contains  kinetic energy term, baryon meson interaction term which produce baryon mass, self-interaction of vector mesons which generates the dynamical mass of vector mesons, scalar mesons interactions which induce the spontaneous  breaking of chiral symmetry, and the explicit breaking term of chiral symmetry 
 \cite{papa_nuclei}. Further, using mean field approximation we solve the effective Lagrangian density and through Euler Lagrange equation, we find the coupled equation of motion in terms of non-strange scalar filed $\sigma$,   strange scalar field $\zeta$, scalar isovector field $\delta$ and dilaton field $\chi$. Here, $\sigma$ ($\sim$ $(u \bar{u} + d \bar{d})$) is the non-strange scalar isoscalar field,  $\zeta$ ($\sim$ $s\bar{s}$) is strange scalar isoscalar field,  $\delta$ ($\sim$ $(u\bar{u} - d\bar{d})$) is scalar isovector field and $\chi$ is the  scalar dilaton field. After solving these equations numerically, we calculate the values of the  light quark condensates $\left\langle \bar{q}q\right\rangle$, strange quark condensates $\left\langle \bar{s}s\right\rangle$  and gluon condensates  $\left\langle  \frac{\alpha_{s}}{\pi} {G^a}_{\mu\nu} {G^a}^{\mu\nu}
\right\rangle$, at finite temperature ($T$), baryonic density ($\rho_B$), strangeness fraction ($f_s$) and isospin asymmetric parameter $I$  \cite{rahul2,rahul2}. Also, the strangeness fraction $(f_s)$ and isospin asymmetric parameter $(I)$ are defined as $f_s$ = $\frac{{\sum}_i |s_i| \rho_i}{\rho_B}$ and $I$ = -$\sum_i \frac{I_{3i}\rho_i}{2\rho_B}$, respectively. Where, $s_i$ is the number of strange quarks of baryons, $\rho_i$ is the number density of the baryon of $i^{th}$ type and $I_{3i}$ is the z-component of the isospin for the $i^{th}$ baryons \cite{papa_nuclei}.

 Then we considered these in-medium values of various condensates as input in the QCD sum rules analysis. In QCD sum rules we start with two point correlation function, which is the Fourier transform of 
 the expectation value of time-ordered product of isospin averaged current of corresponding meson.  Further, this two point function is decomposed into vacuum and nucleon dependent part. Furthermore, the correlation function $T_{\mu \nu}^N (\omega,q)$ that appears in the nucleon dependent part of two point correlation function can be expressed in terms of forward scattering amplitudes $T_N(\omega,q)$ for vector mesons \cite{rahul1} and $T^0 _N(\omega,q)$ for pseudoscalar open charm meson \cite{rahul2,rahul3}.  Then the forward scattering amplitude in limit of $q$ $\to$ 0, can be expressed in terms of spin averaged spectral density, which can be further parametrised in terms of three unknown parameters $a$, $b$ and $c$. 
 Moreover, the in-medium mass of vector and psedoscalar meson is defined as  
   $
\delta m_{D} = 2\pi \frac{m_N + m_{D}}{m_N m_{D}} \rho_N a_{D}$ \cite{rahul1, wang2}.
To evaluate the in-medium mass, the unknown parameter $a$ has be to be determined. The general criteria in QCD sum rule approach equate the Borel transformation of the scattering matrix on the phenomenological side with the  Borel transformation of the scattering matrix
for the OPE side (which contains various condensates). The detailed procedures can be found in ref. \cite{rahul1, rahul2, rahul3, wang2}. This procedure will lead us to one equation with two unknown parameters. To solve that equation we differentiate it w.r.t. $\frac{1}{M^2}$, and then these two equations can be solved to find two unknown parameters.  Using this criteria we can find the in-medium mass of open charm meson for  in-medium masses of vector $D^*$, psuedoscalar $D$, strange pseudoscalar $D_s$ mesons.  

\section{$^3 P_0$ Model and In-medium Strong Decays of $Y(4008)$ and $\psi(4040)$}
To calculate the in-medium partial decay width of $Y(4008)$ and $\psi(4040)$ mesons, we use $^3 P_0$ model. In this model, the quark and antiquark pair is supposed to be created with the vacuum $0^{++}$ quantum numbers \cite{micu,yao}.  This model had been used in the past to evaluate decay widths of OZI allowed two body decays mesons and baryons \cite{ackleh,close,barn1,barn2}. In order to investigate the in-medium partial decay widths of above mentioned mesons 
we use the transition operator as taken in \cite{is}, and find the helicity amplitude (for general decay $Y \to D\bar{D}$) given by \cite{liu} 
\begin{widetext}
\begin{align}\label{eq:M}
 \mathcal{M}^{M_{J_{Y} } M_{J_{D} } M_{J_{\bar{D}}}} = \gamma  \sqrt {8E_{Y} E_{D} E_{\bar{D}} } \sum_{\substack{M_{L_{Y} } ,M_{S_{Y} } ,M_{L_{D} }, \\M_{S_{D} } ,M_{L_{\bar{D}}} ,M_{S_{\bar{D}} } ,m} }\langle {1m;1 - m}|{00} \rangle \nonumber \\
 \times \langle {L_{Y} M_{L_{Y} } S_{Y} M_{S_{Y} } }| {J_{Y} M_{J_{Y} } }\rangle \langle L_{D} M_{L_{D} } S_{D} M_{S_{D} }|J_{D} M_{J_{D} } \rangle\langle L_{\bar{D}} M_{L_{\bar{D}} } S_{\bar{D}} M_{S_{\bar{D}} }|J_{\bar{D}} M_{J_{\bar{D}} }\rangle \nonumber \\
  \times\langle\varphi _{D}^{13} \varphi _{\bar{D}}^{24}|\varphi _{Y}^{12}\varphi _0^{34} \rangle
\langle \chi _{S_{D} M_{S_{D} }}^{13} \chi _{S_{\bar{D}} M_{S_{\bar{D}} } }^{24}|\chi _{S_{Y} M_{S_{Y} } }^{12} \chi _{1 - m}^{34}\rangle I_{M_{L_{D} } ,M_{L_{\bar{D}} } }^{M_{L_{Y}} ,m} (\textbf{K}).
\end{align}
\end{widetext}
Here in the above equations, $E_{Y}$= $m_{Y}$, $E_{D}$ = $\sqrt{m_{D}^{*2} + K_{D}^2}$ and $E_{\bar{D}}$ = $\sqrt{m_{\bar{D}}^{*2} + K_{\bar{D}}^2}$ represent the energies of mesons. Further, $m_{Y}$ is the  mass of parent meson whereas, $m^*_{D}$ and $m^*_{\bar{D}}$ represented the medium masses of $D$ and $\bar{D}$ meson respectively.  Also, $\gamma$ is the strength of the pair creation in the vacuum and its value is taken to  be 8.42 \cite{is}.
 We then calculate the spin matrix elements $\langle  \chi _{S_{D} M_{S_{D} }}^{13} \chi _{S_{\bar{D}} M_{S_{\bar{D}} } }^{24}|\chi _{S_{Y} M_{S_{Y} } }^{12} \chi _{1 - m}^{34}\rangle$ in terms of the Wigner's 9j symbol and the flavor matrix element $\langle\varphi _{D}^{13} \varphi _{\bar{D}}^{24}|\varphi _{Y}^{12}\varphi _0^{34} \rangle$ in terms of isospin of quarks as done in refs. \cite{is, liu, yao}.
  In \cref{eq:M},
  $I_{M_{L_{D} } ,M_{L_{\bar{D}} } }^{M_{L_{Y}} ,m} (\textbf{K})$  represents the spatial integral and is expressed in terms of
  wave functions of the parent and daughter mesons.
Further, in the present investigation, we use simple harmonic oscillator type wave functions defined by

\begin{align}
\psi_{nL{M_L}}= (-1)^n(-\iota)^L R^{L+\frac{3}{2}} \sqrt{\frac{2n!}{\Gamma(n+L+\frac{3}{2})}} \exp\Big{(}\frac{-R^2k^2}{2}\Big{)} L_n^{L+\frac{1}{2}}(R^2k^2) Y_{lm}(\bf{k}).
\label{eq:wave}
\end{align} 
 Here, $R$ is the radius of the meson,
 $L_n^{L+\frac{1}{2}}(R^2 {k}^2)$  represents associate Laguerre polynomial and 
 $Y_{lm}(\bf {k})$ denotes the spherical harmonic function. The expression for the decay amplitudes for the given decay modes are calculated as
 $\mathcal{M}$($1^- \to 0^- + 0^-$) = $-\frac{\sqrt{3}}{18} \gamma \sqrt{8E_A E_B E_C}I_{00}$,  $\mathcal{M}$($1^- \to 0^- + 1^-$) = $-\frac{\sqrt{6}}{18} \gamma \sqrt{8E_A E_B E_C}I_{00}$  and $\mathcal{M}$($1^- \to 1^- + 1^-$) = $({\frac{\sqrt{5}}{9}-\frac{1}{18}}) \gamma \sqrt{8E_A E_B E_C}I_{00}$. Here the explicit expression of special integral can be given as
 
\begin{align} \label{eq:integ}
I_{00}& =\left(k\frac{exp\left(\frac{k^2}{2}\left(\frac{\left( \frac{R_D^2m_1}{m_1+m_3}+\frac{R_{\bar{D}}^2m_2}{m_2+m_4}\right)^2}{(R_Y^2 + R_D^2 +R_{\bar{D}}^2)}-(\frac{R_Dm_1}{m_1+m_3})^2-(\frac{R_{\bar{D}}m_2}{m_2+m_4})^2\right)\right)}{2\sqrt{5}\pi^{\frac{5}{4}}\left(R_Y^2 + R_D^2 +R_{\bar{D}}^2\right)^{\frac{13}{2}}R_Y^{\frac{-11}{2}}\left(R_D R_{\bar{D}}\right)^{\frac{-3}{2}}}\right) \nonumber\\
&\times\left[4k^4\left(R_D^2\left(\frac{m_1}{m_1+m_3}\right)+R_{\bar{D}}^2\left(\frac{m_2}{m_2+m_4}\right)\right)^4  \left( R_Y^2 + R_D^2 +R_{\bar{D}}^2  - {R_D}^2\left(\frac{m_1}{m_1+m_3}\right)\right. \right. \nonumber\\
&\left.\left.-R_{\bar{D}}^2 \left(\frac{m_2}{m_2+m_4}\right) \right) +20\left(R_Y^2 + R_D^2 +R_{\bar{D}}^2\right)^2 \left( 3\left(R_Y^2 + R_D^2 +R_{\bar{D}}^2\right)  - 7{R_D}^2\left(\frac{m_1}{m_1+m_3}\right)\right. \right. \nonumber \\
&\left.\left.-7R_{\bar{D}}^2 \left(\frac{m_2}{m_2+m_4}\right) \right)+8k^2\left(R_D^2\left(\frac{m_1}{m_1+m_3}\right)+R_{\bar{D}}^2 \left(\frac{m_2}{m_2+m_4}\right)\right)^2\left(R_Y^2 + R_D^2 +R_{\bar{D}}^2\right)\right. \nonumber \\
&\times\left.  \left( 5\left(R_Y^2 + R_D^2 +R_{\bar{D}}^2\right)  - 7R_D^2\left(\frac{m_1}{m_1+m_3}\right)-7R_{\bar{D}}^2 \left(\frac{m_2}{m_2+m_4}\right) \right)+20\frac{\left(R_Y^2 + R_D^2 +R_{\bar{D}}^2\right)^2}{R_Y^2}\right.\nonumber \\
&\times\left. \left\lbrace k^2\left(R_D^2\left(\frac{m_1}{m_1+m_3}\right)+R_{\bar{D}}^2 \left(\frac{m_2}{m_2+m_4}\right)\right)^2 \left(R_D^2\left(\frac{m_1}{m_1+m_3}\right)+R_{\bar{D}}^2 \left(\frac{m_2}{m_2+m_4}\right)\right. \right. \right.\nonumber \\
&\left.\left.\left. -R_Y^2 - R_D^2 -R_{\bar{D}}^2\right) \right. \right. 
\left. \left.+\left(5\left(R_D^2\left(\frac{m_1}{m_1+m_3}\right)+R_{\bar{D}}^2 \left(\frac{m_2}{m_2+m_4}\right)\right)-3\left(R_Y^2 + R_D^2 +R_{\bar{D}}^2 \right)\right\rbrace\nonumber \right. \right. \nonumber \\
&\left.\left.  \times\left(R_Y^2 + R_D^2 +R_{\bar{D}}^2\right)\right)  +15 \frac{\left(R_Y^2 + R_D^2 +R_{\bar{D}}^2 \right)^4}{R_Y^4}\left(R_Y^2 + R_D^2 +R_{\bar{D}}^2 -R_D^2\left(\frac{m_1}{m_1+m_3}\right)\right.\right. \nonumber \\
&\left.\left. -R_{\bar{D}}^2 \left(\frac{m_2}{m_2+m_4}\right)\right) \right].
\end{align} 

 
%
Furthermore, using the Jacob-Wick formula \cite{is,JW}, the  helicity amplitude can be transformed  into the partial wave amplitude:
 
 \begin{align}
\mathcal{M}^{JL} (Y \to D \bar{D}) = \frac{{\sqrt {2L + 1} }}{{2J_Y  + 1}}\sum{M_{J_D } ,M_{J_{\bar{D}} } } \langle {L0JM_{J_Y } } |{J_Y M_{J_Y } }\rangle \nonumber \\
\times \left\langle {J_D M_{J_D } J_{\bar{D}} M_{J_{\bar{D}} } } \right|\left. {JM_{J_Y } } \right\rangle M^{M_{J_Y } M_{J_D } M_{J_{\bar{D}} } } (\vec{K}),
\label{eq:MG}
\end{align}
 
where $\vec{J}=\vec{J}_D+\vec{J}_{{\bar{D}}}$, $\vec{J_Y}=\vec{J_D}+\vec{J}_{\bar{D}}+\vec{L}$, $M_{J_Y}=M_{J_D}+M_{J_{\bar{D}}}$. Also the expression of decay width is given as
\begin{align}
\Gamma  = \pi^2 \frac{|{\vec{K}}|}{m_{Y}^2}\sum_{JL} |{\mathcal{M}^{JL}}|^2,
\label{eq:G}
\end{align}

where, $|\vec{K}|$ represents the momentum of the $D$ and $\bar{D}$ mesons in the rest mass frame of $Y$ meson and is given by, 
\begin{align}
 |\vec{K}|= \frac{{\sqrt {[m_{Y}^{2}  - (m^*_D  - m^*_{\bar{D}} )^2 ][m_{Y}^{2}  - (m^*_D  + m^*_{\bar{D}} )^2 ]} }}{{2m_{Y} }}.
\label{eq:K}
\end{align}
Here, for the decay $\psi(4040) \to D\bar{D}$ the values corresponding to $Y$ will be replaced by $\psi(4040)$. Thus, through the in-medium masses of above mentioned mesons we can calculate the in-medium partial decay widths of above mentioned processes.  
%
\section{Results and discussion}
We shall now describe the various parameters used in the present analysis and elaborate the results of the in-medium partial decay width of the processes $\psi$(4040)/Y(4008) $\to$ $D\bar{D}$, $\psi$(4040)/Y(4008) $\to$ $D^*\bar{D}^*$, $\psi$(4040)/Y(4008) $\to$ $D\bar{D}^*$ and $\psi$(4040)/Y(4008) $\to$ $D_s\bar{D_s}$.  We take the  masses of $u$, $d$, $s$ and $c$ quaks as 0.33, 0.33, 0.55 and 1.6 GeV respectively \cite{is,close2}.    
 Further the value of $R_A$ is chosen to be 3.13 and 2.27 GeV$^{-1}$ for $\psi$(4040) and $Y(4008)$ states respectively \cite{is}.
 
 \begin{figure}
\centering
\includegraphics[width=14cm,,height=12cm]{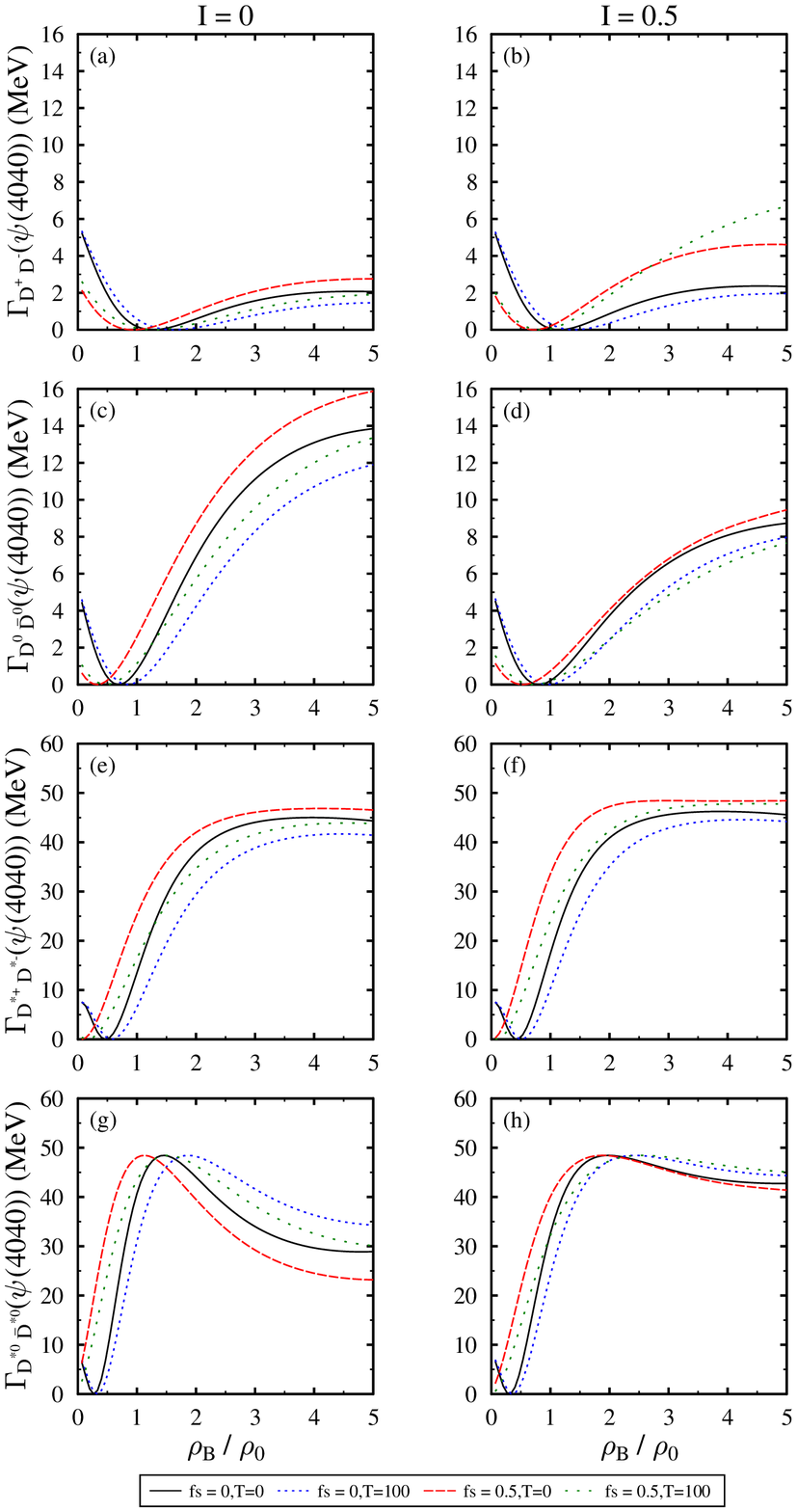}
\caption{Figure shows the variation of in-medium partial decay widths  of the decays $\psi(4040) \to D^+D^-$/$\psi(4040) \to D^0 \bar{D}^0$ and $\psi(4040) \to D^{*+}D^{*-}$/$\psi(4040) \to D^{*0} \bar{D}^{*0}$ as a function of baryonic density in an isospin asymmetric strange hadronic medium.}\label{fig:f1}
\end{figure}

 \begin{figure}
\centering
\includegraphics[width=14cm,,height=12cm]{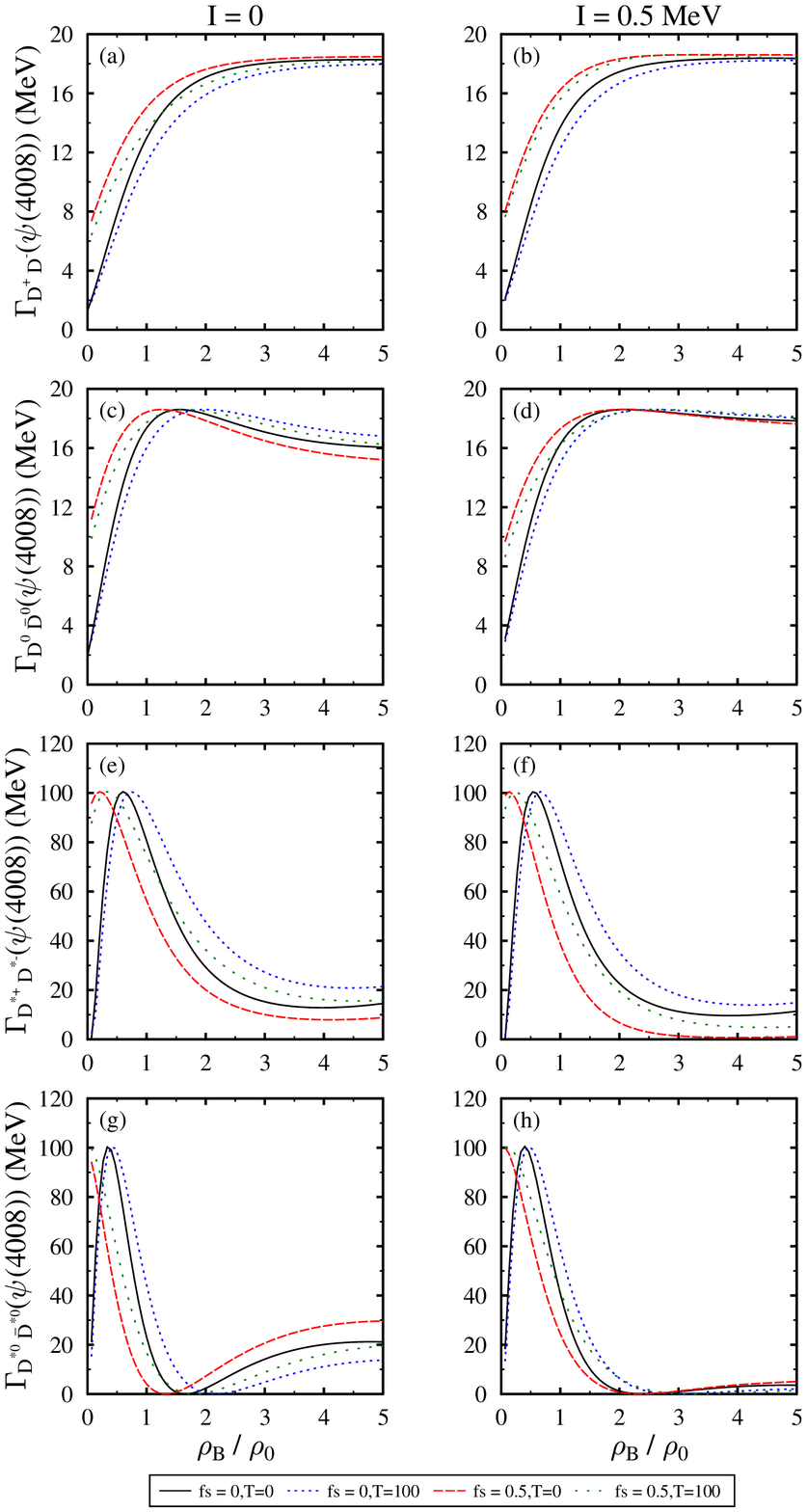}
\caption{Figure shows the variation of in-medium partial decay widths  of the decays $Y(4008) \to D^+D^-$/$Y(4008) \to D^0 \bar{D}^0$ and $Y(4008) \to D^{*+}D^{*-}$/$Y(4008) \to D^{*0} \bar{D}^{*0}$ as a function of baryonic density in an isospin asymmetric strange hadronic medium.}\label{fig:f2}
\end{figure}

 \begin{figure}
\centering
\includegraphics[width=14cm,,height=12cm]{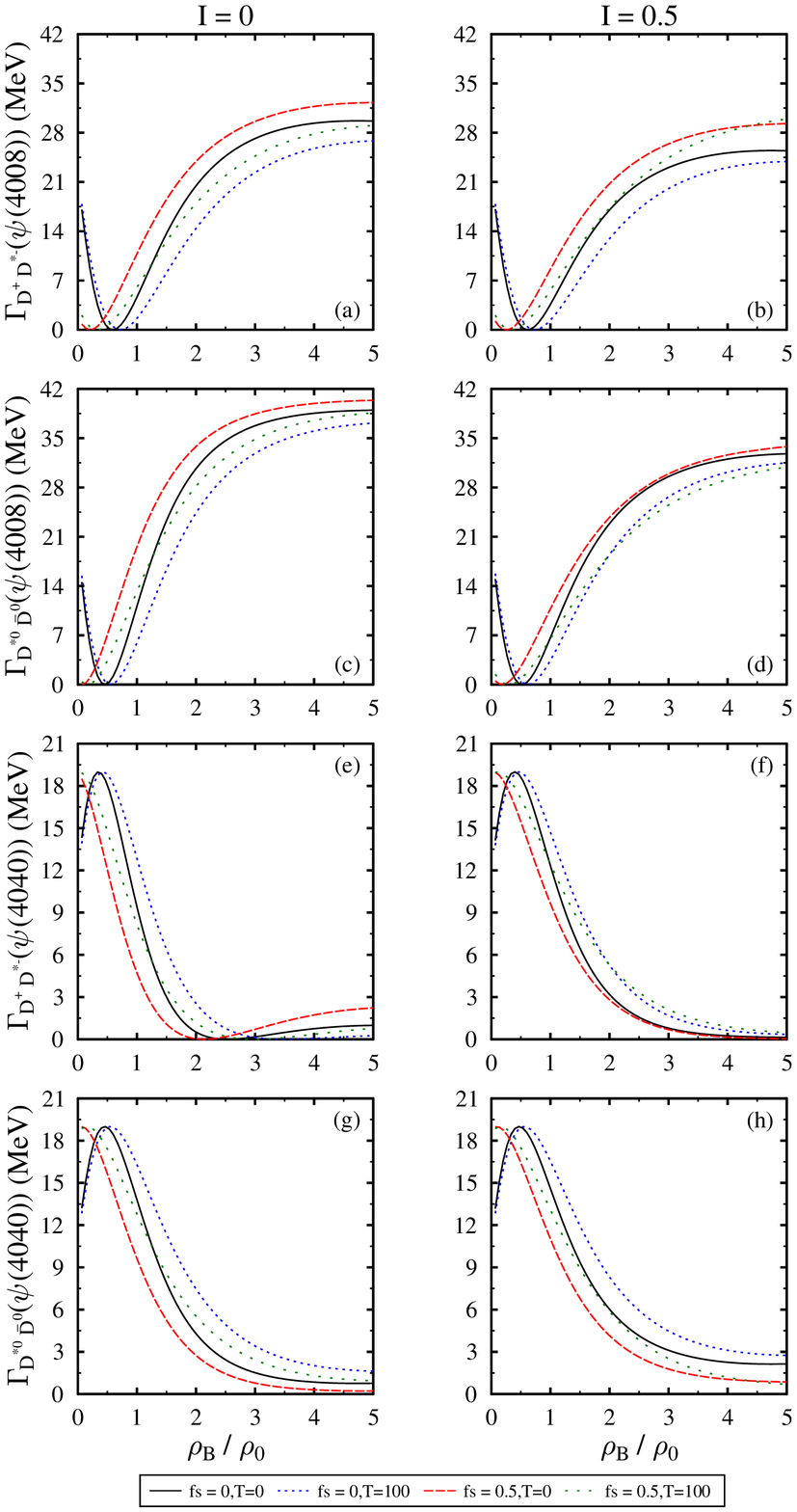}
\caption{Figure shows the variation of in-medium partial decay widths  of the decays $Y(4008) \to D^+D^{*0}$/$Y(4008) \to D^{*0} \bar{D}^0$ and $\psi(4040) \to D^+D^{*0}$/$\psi(4040) \to D^{*0} \bar{D}^0$ as a function of baryonic density in an isospin asymmetric strange hadronic medium.}\label{fig:f3}
\end{figure}

 \begin{figure}
\centering
\includegraphics[width=14cm,,height=12cm]{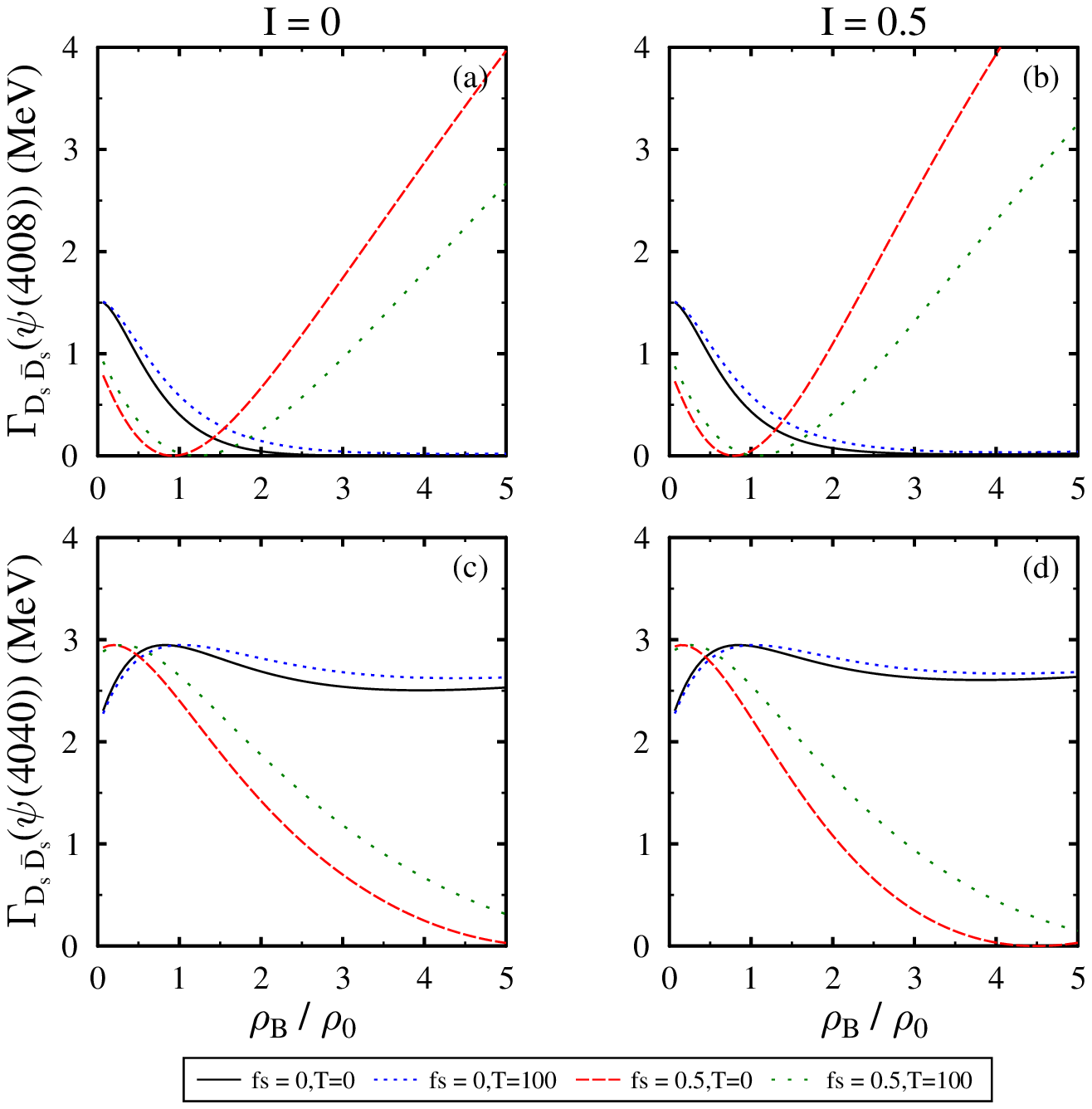}
\caption{Figure shows the variation of in-medium partial decay widths  of the decays $Y(4008) \to D_s \bar{D}_s$ and $\psi(4040) \to D_s \bar{D}_s$ as a function of baryonic density in an isospin asymmetric strange hadronic medium.}\label{fig:f4}
\end{figure}

 In \cref{fig:f1} we present the variation of the partial decay width of the processes $\psi(4040) \to D^+ + D^-$, $\psi(4040) \to D^0 + \bar{D^0}$, $\psi(4040) \to D^{*+} + D^{*-}$ and $\psi(4040) \to D^{*0} + \bar{D^{*0}}$  as a function of baryonic density, in an isospin asymmetric hot and strange hadronic matter. We observe that, for  any constant value of the  temperature ($T$), isospin asymmetric parameter ($I$) and strangeness fraction ($f_s$) of the medium, the value of the partial decay width of the above-mentioned processes first decrease  to  zero till certain value of the baryonic density ($\rho_B$ $\approx$ 0 $-$ 1.5$\rho_0$), then beyond this value it start increasing with an increase in the baryonic density.  The observed node in these partial decay widths is caused by the polynomial part present in \cref{eq:integ}, which becomes zero at this value of baryonic density ($\rho_B$ = 0 $-$ 1.5$\rho_0$). Further, by keeping the other parameters of the medium fixed, on moving from cold ($T = 0$) to hot medium ($T=100$ MeV), the values of the partial decay widths decrease and this happens because of an increase in the mass of the open charm meson as a function of temperature of the medium. Therefore, this causes a drop in the decay channels of the respective decay processes. On the other hand, on shifting from nuclear ($f_s$ = 0) to strange hadronic medium ($f_s$ = 0.5) we observe an enhancement in the value of partial decay width of the respective processes. This can be understood on the basis that mass of the open charm mesons drop more in the strange hadronic medium as compared to nuclear medium and this enhances the decay channel of the processes. Furthermore, the impact of the finite isospin asymmetric parameter ($I$ = 0.5) is to cause different shift in the masses of the isospin doublets ($D^0$, $D^+$), therefore on moving from symmetric ($I = 0$) to isospin asymmetric ($I$=0.5) medium we observe different behaviour of $\Gamma_{D^0{\bar{D}^0}}(\psi(4040))$ and $\Gamma_{D^+{D^-}}(\psi(4040))$.  Here, the observed value of $\Gamma_{D^0{\bar{D}^0}}(\psi(4040))$ is higher as compared to $\Gamma_{D^+{D^-}}(\psi(4040))$.  This can be understood on the basis of the more drop in the mass of $D^0$ meson as compared to $D^+$ meson (for detailed discussion refer \cite{rahul2}).   In addition, in the present investigation, we observe much higher value of strong decay width of $\psi(4040)$ state decaying to vector $D^{*+}{D^{*-}}$ / ${D^{*0}{\bar{D}^{* 0}}}$ pairs as compared to the pseudo-scalar pairs. This behaviour is observed because by using $^3P_0$ model we calculated the decay amplitude (appearing in \cref{eq:MG}) as, $\mathcal{M}$($1^- \to 0^- + 0^-$) = $-\frac{\sqrt{3}}{18} \gamma \sqrt{8E_A E_B E_C}I_{00}$,  and $\mathcal{M}$($1^- \to 1^- + 1^-$) = $({\frac{1}{18}-\frac{\sqrt{5}}{9}}) \gamma \sqrt{8E_A E_B E_C}I_{00}$.  Therefore, due to the dominance of the calculated decay amplitude(for the decay of type $1^- \to 1^-+1^-$) the higher value of in-medium partial decay width is observed. Further, at low baryonic density of the medium the value of $\Gamma_{D^{*0}{\bar{D}^{*0}}}(\psi(4040))$ is more as compared to  $\Gamma_{D^{*+}{{D}^{*-}}}(\psi(4040))$. This is because of the more drop in the mass of $D^{*0}$ meson  as compared to $D^{*+}$. However,  further increase in the baryonic density of the medium causes much big drop in the mass of $D^{*0}$ meson and nodel structure of the wave function comes into play which causes more decrease in the value of $\Gamma_{D^{*0}{\bar{D}^{*0}}}(\psi(4040))$ as compared to $\Gamma_{D^{*+}{{D}^{*-}}}(\psi(4040))$.  
%

In \cref{fig:f2} we show the variation of partial decay widths of the processes $Y(4008) \to D^{+} D^{-}/D^{0} \bar{D^0}$,  in subplot(a - d) and $Y(4008) \to D^{*+} D^{*-} / {D^{*0}{\bar{D}^{* 0}}}$, in subplot(e - h) as a function of baryonic density of the medium. We observe an increase in the value of  $\Gamma_{D^+{D^-}}(\psi(4008))$ and $\Gamma_{D^0{\bar{D}^0}}(\psi(4040))$ with an increase in the  baryonic density of the medium, by keeping the other parameters of the medium fixed. 
 Furthermore, the partial decay width of $Y(4008)$ state decaying to  vector $D^* \bar{D}^*$ pairs first increases till the peak value and then 
 decreases with further increase in the baryonic density of the medium. This happens because  more drop in the mass of $D^*$ mesons cause decrease in the value of polynomial part present in \cref{eq:integ} and this happens because of the nodal structure of the wave functions used in the present investigation. Here we point out that, the different in-medium behaviour of $\Gamma_{D{D}}(\psi(4008))$($\Gamma_{D^*{D^*}}(\psi(4008))$) as compared to $\Gamma_{D{D}}(\psi(4040))$($\Gamma_{D^*{D^*}}(\psi(4040))$) is because of the different masses and $R_A$ values taken for $\psi(4008)$ and $Y(4040)$ states  in $^3  P_0$ model. Further, as mentioned earlier, the large value of $\Gamma_{D{D}}(\psi(4008))$ as compared to  ($\Gamma_{D^*{D^*}}(\psi(4008))$) is because of the dominance of decay amplitude $\mathcal{M}$($1^- \to 1^- + 1^-$) over the decay mode $\mathcal{M}$($1^- \to 0^- + 0^-$). In addition to baryonic density of the medium the finite strangeness fraction, isospin asymmetric parameter and temperature of the medium   also have  significant impact the values of partial decay widths.  For example, on moving from nuclear to strange hadronic medium the values of $\Gamma_{D^+{D^-}}(\psi(4008))$ and $\Gamma_{D^0 \bar{D}^0}(\psi(4008))$ increase. However, beyond the peak value ($\approx$ 17 MeV), the decay channel for the decay $Y(4008)$ $\to$ $D^0 \bar{D}^0$ drops because of the nodal structure that comes into play with the large drop in the mass of the neutral pseudoscalar $D^0$ meson (in strange hadronic matter, as discussed earlier). On the other hand, before the peak value ($\approx$ 100 MeV) values of   $\Gamma_{D^{*+}{D^{*-}}}(\psi(4008))$ and $\Gamma_{D^{*0} \bar{D}^{*0}}(\psi(4008))$ are more than that of non-strange medium, whereas beyond the peak value this trend reverses. This drop in the  partial decay widths occurs because of the  nodal structure of the wave functions. Also, the large drop in the neutral vector $D^{*0}$ meson causes much more drop in the partial decay width of process $Y(4008)$ $\to$ $D^{*0} +  \bar{D}^{*0}$ so that we observed nodes in this particular decay. However, beyond the nodal point, i.e. $\rho_B$ $\approx$ 1.2$\rho_0$  the finite strangeness faction enhances the value of $\Gamma_{D^{*0} \bar{D}^{*0}}(\psi(4008))$. Apart from this, the impact of finite temperature is opposite to the above mentioned trend. For example, in a region where the nodal structure is dominant (i.e. decay width is decreasing) the finite temperature causes increase in the partial decay width, whereas on moving from symmetric to isospin asymmetric medium we observe different modification in the values of  $\Gamma_{D^{*+}{D^{*-}}}(\psi(4008))$ and $\Gamma_{D^{*0} \bar{D}^{*0}}(\psi(4008))$. 

Moreover, in \cref{fig:f3} we represent the values of the partial decay widths of the processes $Y(4008) \to D^+D^{*-}$/ $D^{*0} \bar{D}^0$ and $\psi(4040) \to D^+D^{*-}$/$D^{*0} \bar{D}^0$ as a function of baryonic density of the medium. We observe that, in cold symmetric nuclear medium,  with an increase in the baryonic density of the medium till $\rho_B$ = 0.5$\rho_0$, the values of $\Gamma_{D^{+}{D^{*-}}}(\psi(4008))$/$\Gamma_{D^{0}{\bar{D}^{*0}}}(\psi(4008))$ decrease whereas, the values of $\Gamma_{D^{+}{D^{*-}}}(\psi(4040))$/$\Gamma_{D^{0}{\bar{D}^{*0}}}(\psi(4040))$ increase. However, beyond this point, further increase in baryonic density reverses this trend for the both the processes. 
 This happens because of the polynomial part present in the integration \cref{eq:integ}, which becomes zero at $\rho_B$ = 0.5$\rho_0$ and behaves opposite for $Y(4008)$ and $\psi(4040)$ states.
Further, in cold symmetric medium, on moving from nuclear to strange hadronic medium (i.e., $f_s$ = 0 $\to$ 0.5)  the values of  partial decay widths  $\Gamma_{D^{+}{D^{*-}}}(\psi(4008))$ and $\Gamma_{D^{0}{\bar{D}^{*0}}}(\psi(4008))$  increase and this happens because of the drop in the mass of open charm meson. However,  the values of  $\Gamma_{D^{+}{D^{*-}}}(\psi(4040))$ and $\Gamma_{D^{0}{\bar{D}^{*0}}}(\psi(4040))$ decrease in the presence of strange hyperons (in addition to nucleons) in the medium. On the  other hand, finite temperature ($T=100$ MeV) causes drop in the values of $\Gamma_{D^{+}{D^{*-}}}(\psi(4008))$ and $\Gamma_{D^{0}{\bar{D}^{*0}}}(\psi(4008))$, whereas, an enhancement in the values of $\Gamma_{D^{+}{D^{*-}}}(\psi(4040))$ and $\Gamma_{D^{0}{\bar{D}^{*0}}}(\psi(4040))$, on keeping the other properties of the medium fixed. 
 
  Furthermore, in \cref{fig:f4} we present the values of the partial decay widths of the processes $Y$(4008) $\to$ $D_s \bar{D}_s$ and $\psi$(4040) $\to$ $D_s \bar{D}_s$ as a function of baryonic density in hot and dense asymmetric strange haronic medium. We notice that in cold and symmetric nuclear medium the values of $\Gamma_{D_{s}{\bar{D}_{s}}}(\psi(4040))$ and $\Gamma_{D_{s}{\bar{D}_{s}}}(\psi(4008))$ decrease with an increase in nuclear density of the medium. Finite strangeness fraction causes an increase in the value of $\Gamma_{D_{s}{\bar{D}_{s}}}(\psi(4008))$, whereas drop in the value of $\Gamma_{D_{s}{\bar{D}_{s}}}(\psi(4040))$. These in-medium partial decay widths are much sensitive to the presence of the strange hadrons in the medium, and this can be understood on the basis of the quark content of $D_s$($c \bar{s}$) meson. Further, as discussed in our previous work \cite{rahul3}, the in-medium mass (calculated using QCD sum rules) of $D_s$ meson is much sensitive to the in-medium strange quark condensates $\langle s\bar{s}\rangle$ (calculated using chiral SU(3) model) which is much sensitive to the strange scalar field $\zeta$ and this field is more dependent on the presence of the strange quarks in the medium. In this sense, the significant change in the mass of the $D_s$ meson at finite strangeness fraction causes significant change in the values of $\Gamma_{D_{s}{\bar{D}_{s}}}(\psi(4008))$ and  $\Gamma_{D_{s}{\bar{D}_{s}}}(\psi(4040))$.  
  
     Further, finite temperature of the medium enhances (decreases) the probability of $Y(4040)$ ($\psi(4008)$) state to decay to $D_s \bar{D}_s$ pairs. This happens as the finite temperature causes an increase in the mass of $D_s$ meson and this further causes an increase (decrease) in the values of polynomial part present in \cref{eq:integ} for the decay $Y$(4040) $\to$ $D_s \bar{D}_s$ ($Y$(4008) $\to$ $D_s \bar{D}_s$).       
    Also, because of the more vacuum mass of $\psi$(4040) state as compared to $Y$(4008) state the vacuum value  of $\Gamma_{D_{s}{\bar{D}_{s}}}(\psi(4040))$ is more than the value of $\Gamma_{D_{s}{\bar{D}_{s}}}(\psi(4008))$. 
      However, in future if we can confirm this value for the state $Y(4008)$ in strange hadronic matter, then this may prove the state with quantum number 3$^3 S_1$. However, before to make any definite conclusion more work in the theoretical as well as experimental side is required.  
 
  \begin{table} 
\begin{tabular}{|l|l|l|l|l|}
\hline
 &\multicolumn{2}{c|}{$\rho_0$} & \multicolumn{2}{c|}{$4\rho_0$}   \\ \hline
&$\frac{\Gamma_{D^0 \bar{D}^0}}{\Gamma_{D^{*0} \bar{D}^0+c.c.}}$  & $\frac{\Gamma_{D^{*0} \bar{D}^{*0}}}{\Gamma_{D^{*0} \bar{D}^0+c.c.}}$&$\frac{\Gamma_{D^0 \bar{D}^0}}{\Gamma_{D^{*0} \bar{D}^0+c.c.}}$&$\frac{\Gamma_{D^{*0} \bar{D}^{*0}}}{\Gamma_{D^{*0} \bar{D}^0+c.c.}}$\\ \cline{2-5}
$Y(4008)$&1.56&2.06&0.42&0.52 \\ \hline
$\psi(4040)$ & 0.083&4.2&17.3&39.6 \\  \hline
\end{tabular}

\caption{Table shows the ratio of the relevant strong decay width  of $Y(4008)$ and $\psi(4040)$ states in cold and symmetric nuclear medium.}
\label{table:tb1}
\end{table}

\begin{table}
\begin{tabular}{|l|l|l|l|l|}
\hline
 &\multicolumn{2}{c|}{$\rho_0$} & \multicolumn{2}{c|}{$4\rho_0$}   \\ \hline
&$\frac{\Gamma_{D^+ {D^-}}}{\Gamma_{D^{*+} {D^-}+c.c.}}$  & $\frac{\Gamma_{D^{*+} {D^{*-}}}}{\Gamma_{D^{*+} {D^-}+c.c.}}$&$\frac{\Gamma_{D^+ {D^-}}}{\Gamma_{D^{*+} {D^-}+c.c.}}$&$\frac{\Gamma_{D^{*0} \bar{D}^{*0}}}{\Gamma_{D^{*0} \bar{D}^0+c.c.}}$\\ \cline{2-5}
$Y(4008)$&0.0099&0.786&0.30&6.85 \\ \hline
$\psi(4040)$ & 18.79&115&1.14&0.75 \\  \hline
\end{tabular}
\caption{Table shows the ratio of the relevant partial decay width  of $Y(4008)$ and $\psi(4040)$ states in cold and symmetric nuclear medium.}
\label{table:tb2}
\end{table}

  We shall now compare the results of the present calculation with the previous works. As far as our knowledge regarding the literature is concerned, no in-medium partial decay widths of $\psi(4040)$ and $Y$(4008) were observed. However in \cite{barn2} authors found the vacuum values of   $\Gamma_{D{\bar{D}}}(\psi(4040))$, $\Gamma_{D^*{\bar{D}^*}}(\psi(4040))$, $\Gamma_{D{\bar{D}^*}}(\psi(4040))$  and $\Gamma_{D_s{\bar{D}_s}}(\psi(4040))$ as 0.1, 33,  33 and 7.8 MeV respectively. On the other hand, in \cite{is} the values of partial decay widths of $\psi(4040)$ states decaying to $D\bar{D}$, $D^* \bar{D}^*$, $D \bar{D}^*$ and $D_s \bar{D}_s$ pairs were observed to be 11.11, 23.39,  43.02 and 2.18 MeV, respectively. 
We can compare  these values with the present results  of 
$\Gamma_{D^{+}{D^{-}}}(\psi(4040))$, $\Gamma_{D^{0}{\bar{D}^{0}}}(\psi(4040))$, $\Gamma_{D^{*+}{D^{*-}}}(\psi(4040))$, $\Gamma_{D^{0*}{\bar{D}^{0*}}}(\psi(4040))$, $\Gamma_{D^{+}{D^{*-}}}(\psi(4040))$, $\Gamma_{D^{0}{\bar{D}^{0*}}}(\psi(4040))$ and $\Gamma_{D_s{\bar{D}_s}}(\psi(4040))$ as 3, 2.3, 13.5, 40, 17, 9.4 and 3 MeV respectively, at nuclear saturation density, zero temperature and symmetric nuclear medium. Likewise, at $\rho_B$ $=$ 4$\rho_0$ the above values shift to 8, 67, 45, 30, 6.6, 0.7 and 2.5 MeV respectively at zero temperature symmetric nuclear medium. Apart from this, in \cref{table:tb1} and \cref{table:tb2} we give numerical results of the  relevant ratios of the partial decay widths. We can compare these results with the results of ref. \cite{is}, where authors found the values (for parent meson  $\psi$(4040))  $\frac{\Gamma_{D \bar{D}}}{\Gamma_{D^{*} \bar{D}+c.c.}}$ and $\frac{\Gamma_{D^* \bar{D}^*}}{\Gamma_{D^{*} \bar{D}+c.c.}}$  as 0.26 and 0.54 respectively. Further, for parent meson $Y(4008)$ the above values were observed to be  1.26 and 6.06 rexpectively. On the other hand,  in ref. \cite{barn2} for parent meson $\psi(4040)$ the values  of  $\frac{\Gamma_{D \bar{D}}}{\Gamma_{D^{*} \bar{D}+c.c.}}$ and $\frac{\Gamma_{D^* \bar{D}^*}}{\Gamma_{D^{*} \bar{D}+c.c.}}$  were observed to be 0.003 and 1 respectively. Here as mentioned earlier, in  \cite{is,barn2}  no medium effects were included. 

On the other hand, in the present analysis, the parent mesons are not subjected to the medium modifications. To the best of our knowledge regarding the literature, no work is available to calculate the in-medium mass of $\psi(4040)$ and $Y(4008)$ states. However, to understand the extent of  medium shift  of parent mesons on the observed partial decay widths, we recall the in-medium shift in the  mass  of $J/\psi$, $\psi(3686)$ and $\psi(3770)$ states calculated using perturbative QCD approach \cite{ko2003} and chiral SU(4) model \citep{su4}. In ref. \cite{ko2003} authors observed that, using the $D$ meson loop effect, the mass of $J/\psi$, $\psi(3686)$ and $\psi(3770)$ states decrease by 0.25$\%$, 3.5$\%$ and    3.3$\%$ from their vacuum values at nuclear saturation density in cold symmetric nuclear medium. However, in ref. \cite{su4} using chiral SU(4) model authors found the decrease in the mass of $J/\psi$, $\psi(3686)$ and $\psi(3770)$ states by 0.27$\%$, 3.17$\%$ and 4.1$\%$ from their vacuum values, at $\rho_B$ = $\rho_0$ cold symmetric nuclear medium.  In the similar way, we expect drop in the masses of $\psi(4040)$ and $Y(4008)$ states. Therefore, at $\rho_B$ = $\rho_0$, if we allow 4$\%$ drop in the masses of  $\psi(4040)$ and $Y(4008)$ states then the values $\Gamma_{D^{+}{D^{-}}}(\psi(4040))$ ($\Gamma_{D^{+}{D^{-}}}(\psi(4008))$), $\Gamma_{D^{0}{\bar{D}^{0}}}(\psi(4040))$ ($\Gamma_{D^{0}{\bar{D}^{0}}}(\psi(4008))$), $\Gamma_{D^{*+}{D^{*-}}}(\psi(4040))$ ($\Gamma_{D^{*+}{D^{*-}}}(\psi(4008))$) and $\Gamma_{D^{0}{\bar{D}^{0}}}(\psi(4040))$ ($\Gamma_{D^{0}{\bar{D}^{0}}}(\psi(4008))$) are observed to  be 0.65 (4), 0.21 (9.7), 13 (79) and 40 (44) MeV, respectively, at $\rho_B$ $=$ $\rho_0$, cold symmetric nuclear medium. Clearly, these values are different from the previously discussed values (without considering the in-medium parent meson masses).  Therefore, before to make definite conclusion about $Y$(4008) state the  prior knowledge of the shift in mass is important and we leave this for the future as it requires separate study. Also, we leave the results of the present investigation wait for a check in the forthcoming future experiments and predictions from the other theoretical models.

The detailed study of mass modification of $\psi(4040)$ and $Y(4008)$ states in hot and dense strange hadronic medium and its possible impact on the in-medium study of the partial decay widths as well as on the possibility of the molecular state will be the aim of our future work.

 \section{Summary}
 \label{summary} 
 Under $^3P_0$ model calculations, the in-medium partial decay widths of $\psi(4040)$ and $Y(4008)$ states decaying to a pair of open charm mesons are studied for the first time.  We observed significant impact of the in-medium mass modifications of $D$, $D^*$ and $D_s$ mesons on the partial decay widths of the processes $\psi(4040)$ and $Y(4008)$ states decaying to $D\bar{D}$, $D\bar{D^*}$, $D^* \bar{D^*}$ and $D_s \bar{D}_s$ pairs. The in medium masses of these open charmed mesons are taken as input in the $^3P_0$ model in-order to observe the in-medium partial decay widths of above mentioned processes. We observed that simple drop in the mass of daughter open charm meson does not simply enhance the decay channel, the nodal structure of the wave function also contribute significantly. Even though the masses of two parent mesons are very close but we observed different variation of their partial decay widths at finite baryonic density, strangeness fractions, temperature and isospin asymmetric parameter of the medium. Here we seek for the more theoretical work from other models for the cross checking of our results of the present analysis.

\acknowledgements
 The author gratefully thanks Dr. Arvind Kumar, Assistant professor at NIT Jalandhar for useful discussion.

\end{document}